# Variability of Radio Signal Attenuation by Single Deciduous Tree Versus Reception Angle at 80 GHz

Jarosław Wojtuń[1], Cezary Ziółkowski[1], Jan M. Kelner[1], Tomas Mikulasek[2], Radek Zavorka[2], Jiri Blumenstein[2], Petr Horký[2], Aleš Prokeš[2], Aniruddha Chandra[3], Niraj Narayan[3], Anirban Ghosh[4]

[1] Institute of Communications Systems, Faculty of Electronics, Military University of Technology, Warsaw, Poland,
jaroslaw.wojtun@wat.edu.pl
[2] Department of Radio Electronics, Brno University of Technology, Brno, Czech Republic
[3] ECE Department, National Institute of Technology, Durgapur, India
[4] ECE Department, SRM University AP, 522240 India

*Abstract*—Vegetation significantly affects radio signal attenuation, influenced by factors such as signal frequency, plant species, and foliage density. Existing attenuation models typically address specific scenarios, like single trees, rows of trees, or green spaces, with the ITU-R P.833 recommendation being a widely recognized standard. Most assessments for single trees focus on the primary radiation direction of the transmitting antenna. This paper introduces a novel approach to evaluating radio signal scattering by a single deciduous tree. Through measurements at 80 GHz and a bandwidth of approximately 2 GHz, we analyze how total signal attenuation varies with the reception angle relative to the transmitter-tree axis. The findings from various directional measurements contribute to a comprehensive attenuation model applicable to any reception angle and also highlight the impact of bandwidth on the received signal level.

*Index Terms*— measurements, path loss model, propagation, single tree, vegetation.

## I. Introduction

The development of mobile networks and the increasingly high requirements imposed on them necessitate the use of increasingly larger radio resources [1]. In fifth-generation (5G) and beyond networks, the potential of using millimeter waves (mmWaves) (i.e., FR2, 24 − 90 GHz, formally 30 − 300 GHz) has been recognized [2]. The mmWave frequency band offers advantages such as high data rates [3] and a low level of interference [4, 5]. On the other hand, the use of mmWave frequencies in wireless communication systems has to overcome challenges such as severe path loss, high atmospheric absorption [6], sensitivity to precipitation (such as rain, snow, and fog) [7, 8], and signal blocking [9], especially in non-line-of-sight (NLOS) conditions.

Implementing new wireless communication systems in the mmWave range significantly expands the possibilities and diversity of telecommunications services, but it requires a thorough understanding of the environment's transmission properties. The propagation characteristics of mmWave channels have been extensively studied in recent years [10, 11]. Numerous measurement campaigns have been conducted to investigate the propagation characteristics of mmWave channels in various scenarios, including indoor [12], outdoor [13, 14], and vehicular [15, 16].

Electromagnetic wave attenuation is a crucial characteristic that influences both the spatial design of wireless networks and the assessment of received signal quality in the presence of environmental interference. This characteristic depends on a wide range of parameters related to the environment – such as landform and cover, as well as the type, shape, and size of obstacles – and the transmission technology used, including frequency range, transmitted power, and antenna systems. The phenomenon of electromagnetic wave attenuation and scattering caused by a single deciduous tree represents one of the simplest propagation scenarios encountered in suburban areas. Despite its spatial simplicity, this scenario has significant implications for millimeter wave propagation.

Many studies have examined the influence of trees on signal attenuation. In works [17–20], researchers analyzed the attenuation caused by trees both with and without leaves. They also investigated the effects of antenna polarization, line-of-sight (LOS), and NLOS conditions, as well as the dependence of attenuation on different parts of the tree crown.

Consequently, the ITU-R P.833 recommendation [21] defines a propagation model that describes how attenuation varies with distance, dimensions, vegetation type, and the electrical parameters of obstacles. However, this recommended model has several limitations related to the positioning of the transmitter (TX), the single deciduous tree, and the receiver (RX), which together define a straight-line propagation path. This scenario does not allow for evaluating scattering effectiveness in NLOS conditions when using a single deciduous tree as a scattering element (broken propagation path). Additionally, the model does not account for the influence of the transmitted signal bandwidth, $B$, on the attenuation value.

In this paper, we propose an extension of the scenario to include cases with varying scattering directions for wave propagation in the 80 GHz range. Additionally, the results consider the impact of the transmitted signal's bandwidth on the attenuation caused by the single deciduous tree.

The relationship between the propagation path loss (PL) attenuation, the reception angle, $\alpha$, and the signal bandwidth was established based on an analysis of measurement data. This data was obtained from a dedicated measurement testbed

This research was funded in part by the National Science Center (NCN), Poland, grant no. 2021/43/I/ST7/03294 (MubaMilWave). For this purpose of Open Access, the author has applied a CC-BY public copyright license to any Author Accepted Manuscript (AAM) version arising from this submission.

used in the experimental scenario. A description of both the scenario and the components of the measurement testbed is provided in Section II. The methodology for determining the propagation path attenuation for various reception angles is presented in Section III. Section IV contains the results of the measurement data analysis, along with the interpolation function that represents the relationship between propagation path attenuation and reception angle while accounting for the transmitted signal's bandwidth. Finally, we summarize the paper.

## II. Measurement Scenario

A detailed description of the measurement scenario can be found in [22]. The measurement campaign was conducted at various times throughout the year in the Orlické Mountains in the Czech Republic.

In this paper, we focus on measurements taken in June 2023. We analyze the scenario involving a single deciduous tree – a willow (Salix caprea) – which stands 4.3 m tall with a diameter of 3 m. Figure 1 illustrates the relative positions of the TX and RX during the measurements, highlighting the angles and distances between them, as well as the distances from both the TX and RX to the tree.

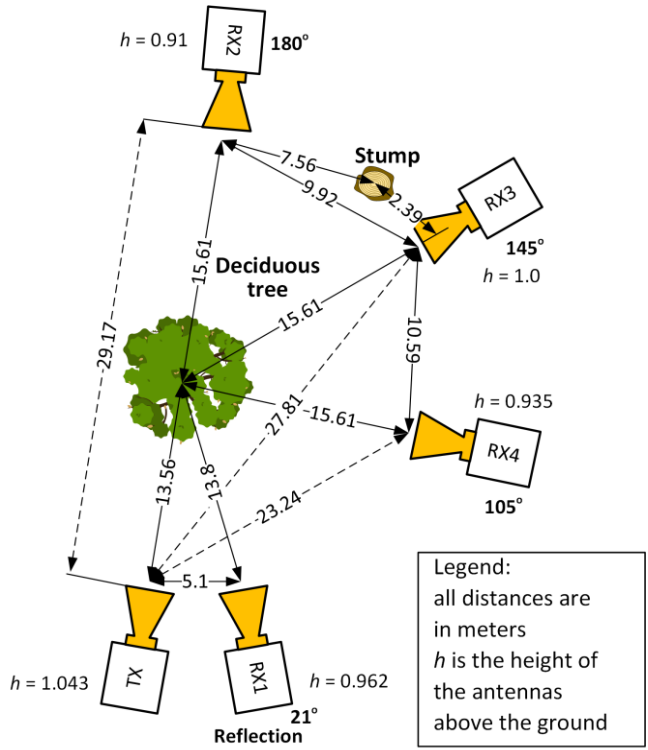

Fig. 1. Measurement scenario.

The measurement testbed consisted of Xilinx Zynq UltraScale+ RFSoC ZCU111 boards, which transmitted the baseband signal and acquired the I and Q components. A simple frequency-modulated continuous wave (FMCW) with a bandwidth of $B = 2.048$ GHz and a $T = 8$ μs duration was used as the baseband signal. Most of the energy of the signal is within the range of ±980 MHz, therefore, the maximum analyse bandwidth is equal 1960 MHz. The signal was up- and down-converted to the mmWave band using the Sivers IMA FC1003E/03 up/down converter. Agilent generators with a 10 MHz rubidium reference clock served as the local oscillators. The signal was transmitted at carrier frequency $f = 81.6$ GHz. Additionally, a power amplifier (Filtronic Cerus 4 AA015), a low-noise amplifier (LNF-LNR55_96WA_SV), and horn or open-ended waveguide antennas (OWGA) were employed.

In [22] the authors provide all details on the calibration of the measurement system, raw data processing, and the methods employed to eliminate crosstalk between the TX and RX antennas.

## III. Path Loss Evaluation versus Reception Angle

To achieve the aim of this paper, i.e., determining how the attenuation of a radio signal varies with the reception angle, we propose a procedure for assessing this attenuation. Our approach involves evaluating the changes in the power level of the received signal in relation to the power level of a reference signal, which is received at an angle of 180°. Additionally, we have considered the bandwidth of the transmitted signal as a relevant parameter in our analysis.

The received signal is processed to estimate the PL. We transform it to the frequency domain (by Fourier transform) and determine its power

$$P_\alpha(m) = \sum_{n=1}^{N}|X_{\alpha,m}(n)|^2, \quad (1)$$

where $X_{\alpha,m}(n)$ is the value of the Fourier transform for a specific spectrum bin, $\alpha$, and analysed impulse, $N$ is the number of the Fourier transform coefficients, which correspond directly with the analysed bandwidth of the signal, and $M$ is the number of analysed impulses.

The mean power for all received impulses is calculated. In our research, we average the power of $M$=1000 impulses

$$\bar{P}_\alpha = \frac{1}{M}\sum_{m=1}^{M} P_\alpha(m). \quad (2)$$

We determine the attenuation correction coefficient, $C_\alpha$, for each reception angle in relation to the direct path, i.e., $\alpha = 180°$.

$$C_\alpha = \frac{\bar{P}_\alpha}{\bar{P}_{\alpha=180°}}. \quad (3)$$

We determine the PL for the direct path.

$$PL_{\alpha=180°}(\text{dB}) = 79 + 7 + 7 + 10\log_{10}(\bar{P}_{\alpha=180°}), \quad (4)$$

where 79 dB is the attenuation of the attenuator that was used in the calibration procedure of the measuring testbed (see [22],

chapter III-B). 7 dBi is a gain of OWGA which was used in this case.

For $\alpha = 105°$ and $145°$, we determine PL as follows

$$PL_\alpha = \frac{PL_{\alpha=180°}}{C_\alpha} \leftrightarrow PL_\alpha(\text{dB}) = PL_{\alpha=180°}(\text{dB}) - 10\log_{10}(C_\alpha). \quad (5)$$

Due to the smaller distance of the receiver from the tree for $\alpha = 21°$ compared to the other cases, we introduce an additional attenuation correction factor $\Delta PL_{\alpha=21°}$ defined below

$$\Delta PL_{\alpha=21°}(\text{dB}) = \text{FSPL}_{\alpha \neq 21°} - \text{FSPL}_{\alpha=21°} = 20\log_{10}(d_{\alpha \neq 21°}) - 20\log_{10}(d_{\alpha=21°}) = 1.07. \quad (6)$$

where FSPL is the free space path loss (FSPL(dB) = $20\log_{10}(d) + 20\log_{10}(f) + 20\log_{10}\left(\frac{4\pi}{c}\right)$), where $d$ represents the distance between the TX and RX antennas, $f$ is the center frequency, and $c$ is the speed of light. $d_{\alpha \neq 21°} = 15.61$ m, $d_{\alpha=21°} = 13.80$ m (see Fig. 1).

Finally, for $\alpha = 21°$, the PL is equal to

$$PL_{\alpha=21°}(\text{dB}) = PL_{\alpha=180°}(\text{dB}) - 10\log_{10}(C_{\alpha=21°}) + \Delta PL_{\alpha=21°}. \quad (7)$$

## IV. PATH LOSS MODEL FOR SELECTED SCENARIO

In this paper, we analyze the selected measurement sub-scenario described in [22], i.e., for a tree with foliage. In this case, for each analyzed receiver position illustrated in Fig. 1, 1000 measurements of the channel impulse response (CIR) were performed. Based on them, we determined the received signal spectra and calculated their powers. The empirical cumulative distribution functions (ECDFs) of the relative received power for the individual receiver positions and maximum bandwidth (i.e., 1960 MHz) are shown in Fig. 2.

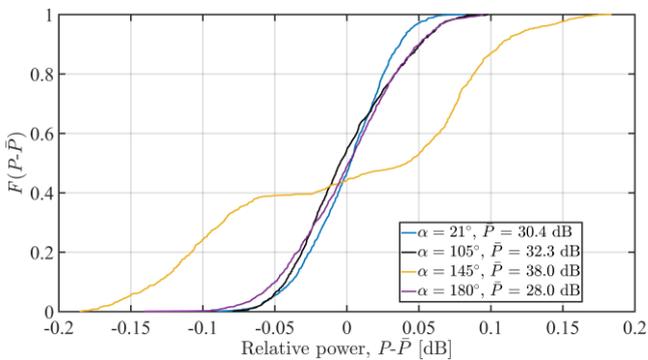

Fig. 2. ECDFs for different receiver positions and maximum bandwidth.

The power distributions for the three receiver positions are close to normal with similar deviations, while for $\alpha = 145°$, the ECDF shape is similar to bimodal. This differentiation may be due to diverse propagation conditions. For the reception angle $\alpha = 145°$, the received signal may be partly scattered by the tree and partly received directly from the transmitter. In this case, the best line-of-sight (LOS) conditions occur, and the average power is the highest ($\bar{P} = 38$ dB, see Fig. 2). For the other reception directions, the signal is primarily scattered by the tree.

Based on the measured powers and considering the reference attenuation of 4.83 dB/m for the analyzed sub-scenario (see [22, Tbl. 3]), we determined the path losses for individual receiver positions, which are depicted in Fig. 3. For the reception angle $\alpha = 21°$, an example PL distribution close to normal is presented.

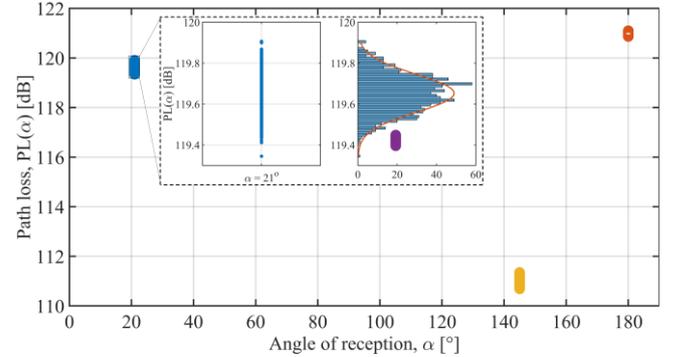

Fig. 3. PLs for different receiver positions and maximum bandwidth.

To determine the continuous PL model as a function of reception angle (for $20° \leq \alpha \leq 180°$), we used third-degree polynomial interpolation, which we can describe as follows:

$$PL(\alpha\;[rad])[\text{dB}] = C\alpha^3 + D\alpha^2 + E\alpha + F, \quad (8)$$

where $C$, $D$, $E$, and $F$ are coefficients of the interpolation polynomial. For the maximum bandwidth, the obtained result is depicted in Fig. 4.

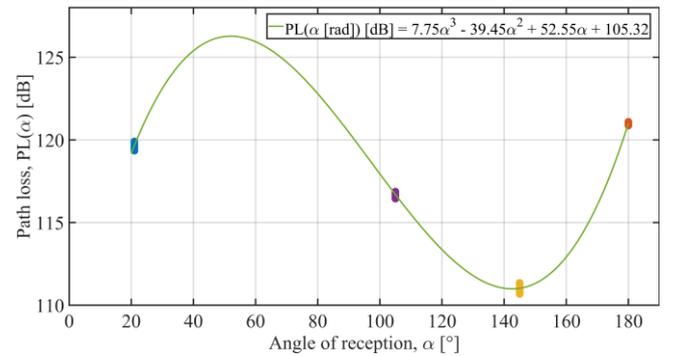

Fig. 4. PL versus reception angle based on third-degree polynomial interpolation for maximum bandwidth.

We can see that the interpolation curve considers the measurement data well. The obtained PL model is in line with intuition. The highest attenuation occurs for the reception direction between 20° and 100°. In these cases, the radio wave scattered by the tree must be reflected either back or sideways. The maximum attenuation occurs in the middle of this range, i.e., around 50°. The smallest attenuation occurs for a reception angle of around 140°. As mentioned above, LOS conditions can partially occur in this case. However, with the increase of the reception angle to $\alpha = 180°$, the attenuation increases again, which is caused by direct scattering introduced by the tree and leaves.

We performed a similar analysis for different bandwidths of the transmitted signal in the range from 200 to 1960 MHz. For this purpose, bandpass filters with the appropriate bandwidth are used to determine the received powers based on the spectra. Next, the procedure described for the maximum bandwidth is repeated for each case. The PL models versus reception angle for different bandwidths $B$ are presented in Fig. 5.

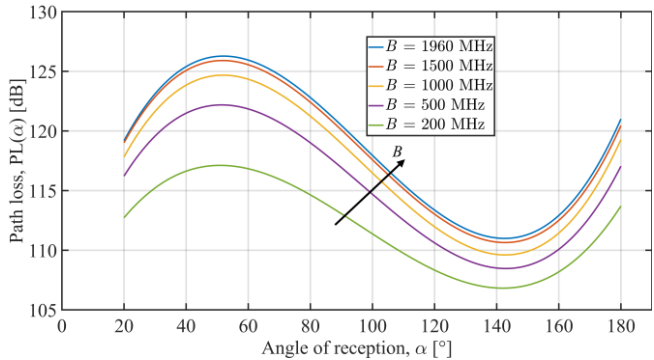

Fig. 5. PL versus reception angle based on third-degree polynomial interpolation for different bandwidths.

We can see that PL increases with increasing bandwidth $B$. The third-degree polynomials described by (8) are again used to model the shapes of the PL curves. The polynomial coefficients of the PL models for the individual bandwidths are given in Table I. The smallest variation in coefficient values occurs for $F$, which means that the attenuation of the received signal that we would obtain near the transmitter is approximately equal to 105 dB and independent of the bandwidth.

TABLE I. POLYNOMIAL COEFFICIENTS FOR PATH LOSS MODEL.

| Bandwidth $B$ [MHz] | Polynomial coefficient | | | |
|---|---|---|---|---|
| | $C$ [dB/rad$^3$] | $D$ [dB/rad$^2$] | $E$ [dB/rad] | $F$ [dB] |
| 200 | 5.14 | –25.95 | 33.98 | 103.81 |
| 500 | 6.63 | –34.28 | 45.25 | 104.31 |
| 1000 | 7.55 | –38.47 | 51.09 | 104.35 |
| 1500 | 7.63 | –38.87 | 51.55 | 105.43 |
| 1960 | 7.75 | –39.45 | 52.55 | 105.32 |

The proposed PL models versus reception angle for different bandwidths constitute a valuable extension of the previously used vegetation attenuation models.

## V. CONCLUSIONS

The propagation attenuation models introduced by vegetation available in the literature consider various scenarios, from a single tree, through a row of trees to an environment such as a park, orchard, or forest. These models usually consider the analysis of attenuation in the vertical plane including the transmitter and the tree(s) as the scattering element. The recommendation [21] introduces, i.a., modeling different components diffracted around the vegetation. However, the receiver is always located in the same plane. In this paper, we propose a novel approach to modeling the scattering introduced by a single tree. The proposed PL model based on empirical measurements and third-degree polynomial interpolation allows us to evaluate the signal attenuation versus reception angle for different bandwidths.

In the near future, we want to extend this modeling method for other weather conditions, i.e. for the remaining sub-scenarios described in [22]. We aim to analyze how bandwidth affects signal attenuation across different seasons. Additionally, we plan to repeat the measurements for various types of trees and evaluate the universality of the developed model.

ACKNOWLEDGMENT

This work was co-funded by the Czech Science Foundation under grant no. 23-04304L, the National Science Centre, Poland, under the OPUS call in the Weave program, under research project no. 2021/43/I/ST7/03294 acronym 'MubaMilWave' and by the Military University of Technology under grant no. UGB/22-478/2024/WAT, and chip-to-startup (C2S) program no. EE-9/2/2021-R&D-E sponsored by MeitY, Government of India.